# The Initial Approximations for Achromatic Doublets of the XVIII Century

Igor Nesterenko
*FRIB/NSCL, Michigan State University, East Lansing, MI 48824, USA*
*Budker Institute of Nuclear Physics, Novosibirsk, 630090, RUSSIA*
(corresponding author, e-mail: nesterenko@frib.msu.edu)

**Abstract**
Analysis of the both type (flint-forward and crown-forward) achromatic doublets was carried out. The investigation revealed possible initial approximations which could be used by opticians at producing of the achromatic doublets in last half of XVIII century. The comparative analysis of approximate versions of achromatic doublets has provided additional explanation to some historical events. One more confirmation that the earliest achromatic doublets were really flint-forward type was found.

**Introduction**
Before manufacturing any optical system, the optician has to perform some calculations. As rule, in XVIII century the "thin-lens" approximation was used. This approximation gives a good agreement with the final result especially at small relative apertures. For achromatic object-lens consisting of two glasses, it is necessary to calculate four radii. This will require a four equations or conditions.
One of them is the focal length of the doublet and other is achromaticity condition. At the "thin-lens" approximation the total optical power $P$ of the doublet is

$$P \equiv \frac{1}{F} = P_f + P_c \qquad (1)$$

Here $P_f$; $P_c$ – optical powers of the flint and crown lenses respectively, herewith $P_f < 0$; $P_c > 0$ and $P > 0$. $F$ is the chosen focal length of the achromatic doublet. The optical powers for flint and crown lenses can be found from formulas:

$$P_f = (n_f - 1)\left[\frac{1}{R_1} - \frac{1}{R_2}\right]$$
$$P_c = (n_c - 1)\left[\frac{1}{R_3} - \frac{1}{R_4}\right] \qquad (2)$$

Here $n_f$; $n_c$ – refractive indexes of flint and crown glasses respectively at green-yellow wavelength; $R_i$ – radii of the surfaces, where $i = 1, 2, 3, 4$ is number of the surface by light path[1].
The doublet will be achromatic if the optical powers will satisfy to next proportions:

$$\frac{P_c}{P_f} = \frac{\theta_c}{\theta_f} = -\frac{\nu_c}{\nu_f} \qquad (3)$$

Here $\theta_f$; $\theta_c$ – refraction angles[2] of flint and crown equivalent prisms respectively on Figure 1, which have equal dispersion, and $\nu_f$; $\nu_c$ – Abbe numbers for glasses of these prisms. The negative sign is required, because the Abbe numbers are positive values herewith always $\nu_c > \nu_f$. The equations (3) take place at the condition when the apex angles of the equivalent prisms – $\alpha_{c,f}$ and theirs refraction angles – $\theta_{c,f}$ should be small enough, so trigonometric functions of these angles can be replaced by angles themselves in the first-order expansion. Thereby the exact Snell's equation $sin(\alpha_{c,f} + \theta_{c,f}) = n_{c,f} sin(\alpha_{c,f})$ for the equivalent prisms is replaced by the approximate expression $\theta_{c,f} \approx (n_{c,f} - 1)\alpha_{c,f}$ . From Figure 1 it is easy to see, that $sin(\alpha_{c,f}) = H/R_{c,f}$ , and $tan(\theta_{c,f}) = H/F_{c,f} = H \cdot P_{c,f}$ . In case of a small angles in the first-order expansion $\alpha_{c,f} \approx H/R_{c,f}$ and $\theta_{c,f} \approx H/F_{c,f} = H \cdot P_{c,f}$, after substitution these expressions in the approximate equation for $\theta_{c,f}$ above, we obtain the formula for the plano-concave or plano-convex thin lens. The general case for an arbitrary lens is considered in Appendix I.

---

[1] For the crown-forward doublet $R_1$, $R_2$ – radii of the crown lens and $R_3$, $R_4$ – radii of the flint lens.
[2] This is the angle between the incident and refracted rays. Refraction angle has positive value if refracted ray will be deflected in clockwise direction relative of the incidence ray or toward to the optical axis of lens.

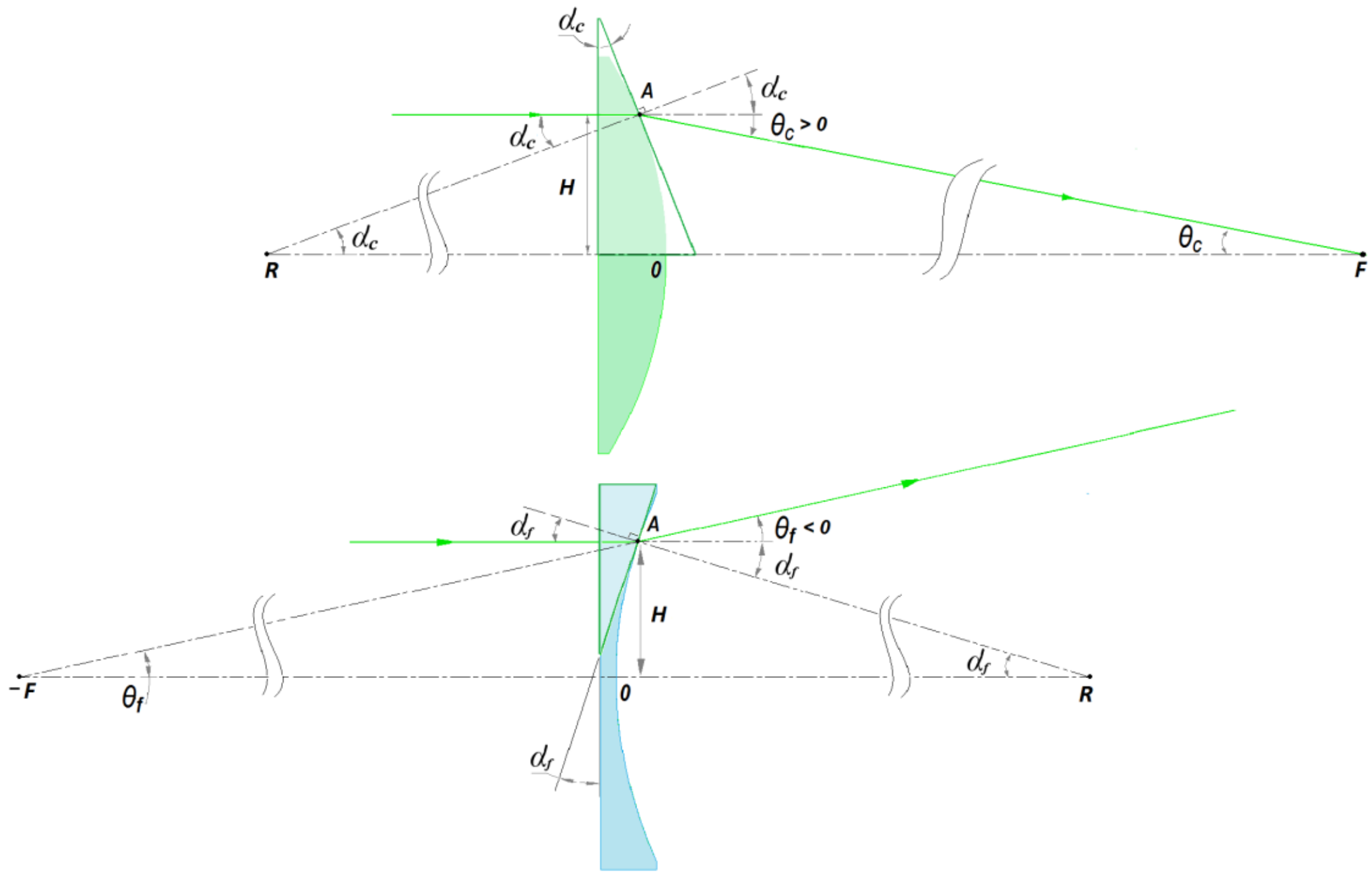

Figure 1. The equivalent prism to the lenses at the given ray height *H*.
*F* – focus position of the lens or it is the focal length relative to point *O; R* – curvature center of the lens surface.

Here we only note, that in the "thin lens" approximation and for case of small angles at retaining only of the first-order in the expansion of trigonometric functions, a spherical aberration does not arise, because the dependence of the optical power from beam height *H* is absent. Therefore, an uncertainty in a choice of correct radii, at which would ensure not only the given focal length and achromaticity of a real doublet but also minimum of spherical aberration, still remains.
In conditions of the XVIII century, optician had a two possible ways for resolving of this uncertainty:

1. The remaining free radii should be set by volitional way. For instance, they may be equal to each other or to be flat or any other preselected value. In addition, the sequence of lenses (flint or crown in front) in the doublet should be chosen.
2. To take into account the following orders expansion of the trigonometric functions as the thickness of a lenses. In this case, the position of focal plane will depend on the height of incident ray (or on the entrance pupil zone). It will give the missing conditions for the radii.

At the volitional assignments for free radii we have to consider all possible variants for searching acceptable, which have a small spherical aberration and therefore they are suitable for making achromatic doublets. Not violating generality of approach, we can take into account only the versions where the free radii are chosen equal to each other and/or are replaced by flat surfaces. In this case, the total quantity of the approximation variants equal to twenty five for flint-forward and the same quantity for crown-forward type, which are realized the achromatic doublets with the chosen focal length.
As noted above, when using "thin lens" approximation together with first-order of expansion for trigonometric functions a spherical aberration is absent. Therefore for selection of the acceptable versions we have to calculate a "real" lenses. For these calculations Zemax software was used. For optician in XVIII century this meant a necessity of their making. Of course, if we will suppose the theory of spherical aberration was not available.
Here under the “real” doublets we understand the object-lenses:

- With a real central thickness such that for a calculated radii at a specific diameter they can be manufactured. This means the lenses have a reasonable and non-zero edge thickness.

- With air gap between flint and crown lenses (if required) such as to prevent their overlapping and at the same time they are placed at minimum distances from each other as it is possible[3].

The comparison of twenty five versions for each type of a doublets with aperture 32mm and focal length 640mm (focal ratio 20) was done. Herewith, the glass pair with optical parameters for flint ($n_e$=1.58228, $\nu_e$=42.12)[4] and for crown ($n_e$=1.52617, $\nu_e$=60.08) was used.

**Flint-forward achromatic doublets**

Possible versions of the flint-forward doublets are shown on Figures 2 and 3. The formulas for calculation of the radii for the each version are collected in Table I.1, in Appendix I. The versions with only a numeric label are the initial variants. The versions on Figure 2 with additional letter to the number differ from the initial (numbered) variants by reverse orientation of the flint and/or crown lenses. Herewith, the absolute values of radii are the same as in corresponding initial variants.
Many from these versions have an evident technical issue: the crown lens rests on the flint lens in one central point. Therefore, it can be tilted during fixation in the objective cell.

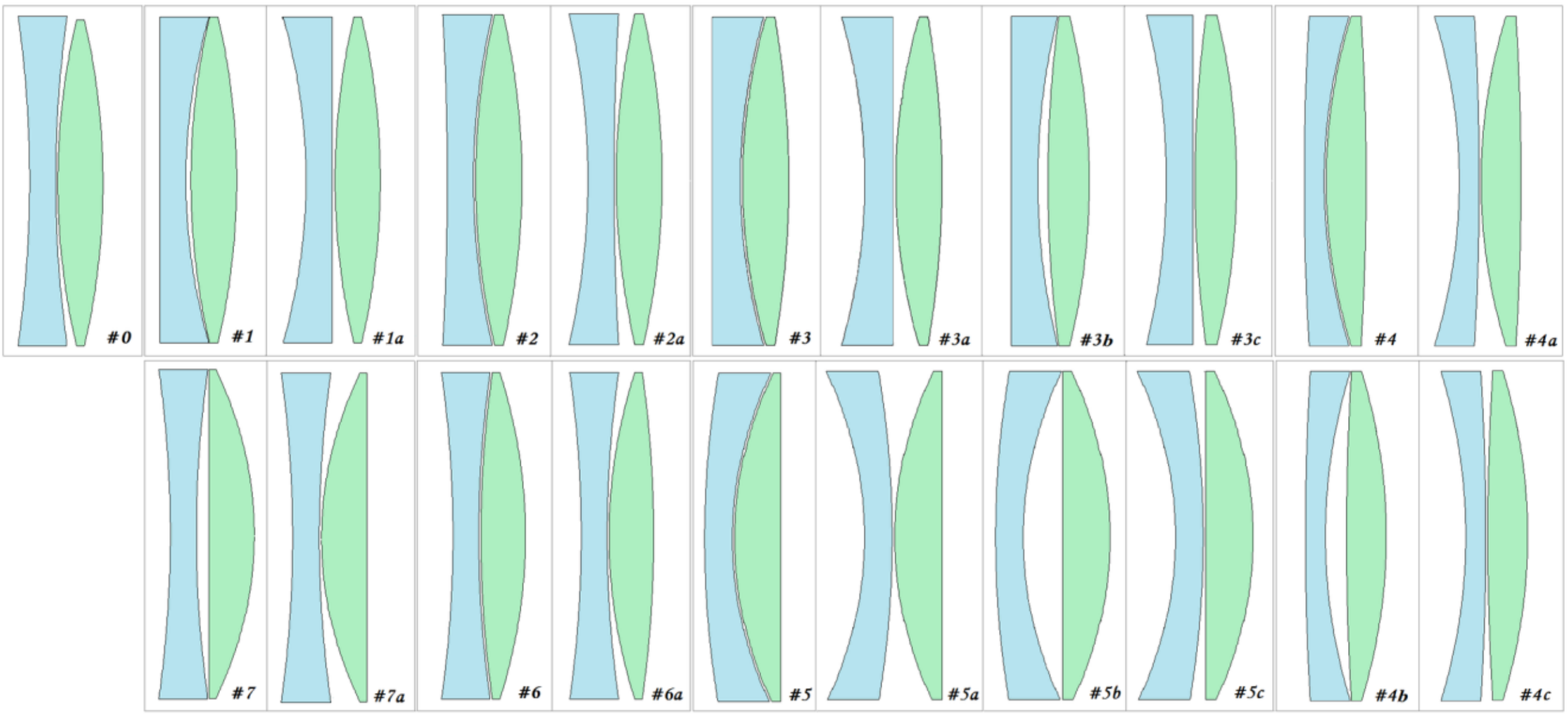

Figure 2. Some approximation versions of the flint-forward achromatic doublets are shown. Light passes through the lenses from left to right, so the blue part is a flint lens. The versions with additional letter in label have the same radii as in the initial versions with corresponding number.

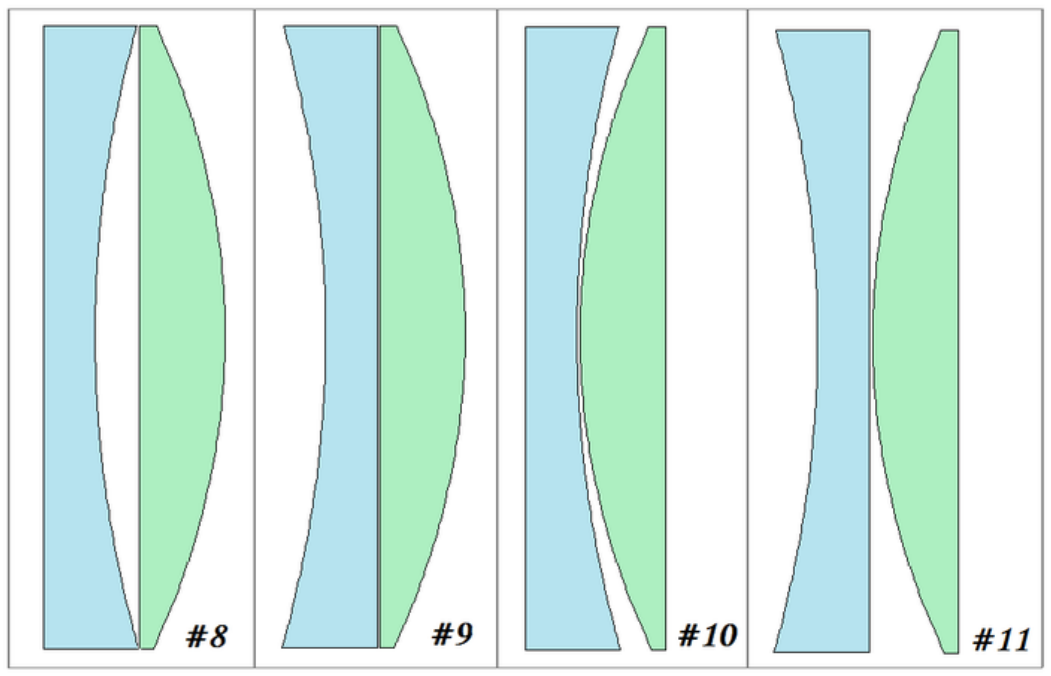

Figure 3. The versions of the flint-forward doublets were attributed to Chester Moor Hall authorship.

Four versions on Figure 3 were described by William Eastland in 1765 during his testimony in the court[5]. Also he referred to the instruction by James Ayscough for making an achromatic lens, who wrote down it from words of Chester Moor Hall. In the court records it was written that both flint and crown lenses have to have a flat surface on one of sides. The other sides should be concave for flint lens and convex for crown lens. Their sequence and relative orientation were not specified. Therefore all four possible versions of the flint-forward doublets were included[6]. Based on the results from Table 1, we can definitely say that from the last four versions only one (#8) could be used as initial approximation for achromatic doublets. The other versions are so far away from the optimum, mainly due to big original spherical aberration. Therefore in practice they hardly were used. At least, their application would hardly led to successful results. The same conclusion can be applied to the versions #6, #7 and almost all variants with the lettered labels.

[3] The achromatic air-spaced doublets outside scope of this article.
[4] The index “e” at refractive index $n_e$ and Abbe number $\nu_e$ means e-line with wavelength 546.1nm.
[5] Reference [3], p.289
[6] Obviously there are the same quantity of crown-forward versions, which will be considered later.

Table 1. Comparison of the versions of the initial approximations for the flint-forward achromatic doublets[7]

| # | $R_1$ | $R_2$ | $R_3$ | $R_4$ | F/D | Peak-to-Valley ($\lambda$=0.546μm) | Strehl Ratio | GEO Radius (μm) | RMS Radius (μm) | TSPH (μm) | TAXC (μm) |
|---|---|---|---|---|---|---|---|---|---|---|---|
| 0 | $-R_2$ | $+R_2$ | $+R_3$ | $-R_3$ | 19.9 | 0.705$\lambda$ | 0.178 | 132.0 | 49.9 | 239.7 | 2.9 |
| 1 | ∞ | $+R_2$ | $+R$ | $-R$ | 19.9 | 0.147$\lambda$ | 0.929 | 26.9 | 10.6 | -47.1 | 3.4 |
| 1a | $-R_2$ | ∞ | $+R$ | $-R$ | 19.7 | 0.941$\lambda$ | ~0.09 | 168.4 | 65.9 | 313.0 | 6.6 |
| 2 | $-R_1$ | $+R$ | $+R$ | $-R$ | 19.9 | 0.285$\lambda$ | 0.752 | 50.3 | 20.5 | 98.8 | 1.9 |
| 2a | $-R$ | $+R_1$ | $+R$ | $-R$ | 19.8 | 0.917$\lambda$ | ~0.09 | 163.4 | 64.7 | 308.4 | 4.8 |
| 3 | ∞ | $+R$ | $+R$ | $-R_4$ | 20.0 | 0.118$\lambda$ | 0.952 | 21.2 | 8.5 | 41.3 | 0.4 |
| 3a | $-R$ | ∞ | $+R$ | $-R_4$ | 19.8 | 1.217$\lambda$ | ~0.12 | 215.8 | 84.9 | 403.3 | 5.4 |
| 3b | ∞ | $+R$ | $+R_4$ | $-R$ | 19.7 | 0.245$\lambda$ | 0.816 | 44.1 | 17.7 | -78.0 | 6.8 |
| 3c | $-R$ | ∞ | $+R_4$ | $-R$ | 19.6 | 0.839$\lambda$ | ~0.09 | 150.5 | 58.9 | 279.4 | 8.4 |
| 4 | $+R_1$ | $+R_2$ | $+R_2$ | $-R_1$ | 20.0 | 0.025$\lambda$ | 0.998 | 5.7 | 1.9 | -8.4 | -0.1 |
| 4a | $-R_2$ | $-R_1$ | $+R_2$ | $-R_1$ | 19.7 | 1.742$\lambda$ | ~0.07 | 305.0 | 119.3 | 561.5 | 7.8 |
| 4b | $+R_1$ | $+R_2$ | $+R_1$ | $-R_2$ | 19.4 | 0.883$\lambda$ | ~0.09 | 157.5 | 63.5 | -279.8 | 15.1 |
| 4c | $-R_2$ | $-R_1$ | $+R_1$ | $-R_2$ | 19.4 | 0.835$\lambda$ | ~0.10 | 150.8 | 58.0 | 272.4 | 14.8 |
| 5 | $+R_1$ | $+R$ | $+R$ | ∞ | 20.0 | 0.073$\lambda$ | 0.982 | 13.9 | 5.3 | -24.7 | 1.4 |
| 5a | $-R$ | $-R_1$ | $+R$ | ∞ | 19.5 | 2.417 $\lambda$ | ~0.06 | 415.4 | 161.6 | 749.3 | 12.0 |
| 5b | $+R_1$ | $+R$ | ∞ | $-R$ | 19.1 | 1.416 $\lambda$ | ~0.12 | 251.4 | 101.2 | -428.5 | 24.8 |
| 5c | $-R$ | $-R_1$ | ∞ | $-R$ | 19.1 | 0.963 $\lambda$ | ~0.09 | 173.2 | 65.9 | 305.6 | 22.8 |
| 6 | $-R$ | $+R$ | $+R$ | $-R_4$ | 19.7 | 0.608$\lambda$ | 0.275 | 107.9 | 43.3 | 207.7 | 5.5 |
| 6a | $-R$ | $+R$ | $+R_4$ | $-R$ | 19.9 | 1.122 $\lambda$ | ~0.11 | 198.3 | 79.3 | 378.8 | 1.3 |
| 7 | $-R$ | $+R$ | ∞ | $-R_4$ | 19.3 | 1.223 $\lambda$ | ~0.13 | 213.1 | 85.2 | 409.4 | 16.1 |
| 7a | $-R$ | $+R$ | $+R_4$ | ∞ | 19.9 | 2.619$\lambda$ | ~0.05 | 451.1 | 180.5 | 868.0 | 0.9 |
| 8 | ∞ | $+R_2$ | ∞ | $-R_4$ | 19.2 | 0.383$\lambda$ | 0.614 | 64.3 | 25.7 | 133.4 | 18.9 |
| 9 | $-R_1$ | ∞ | ∞ | $-R_4$ | 19.3 | 1.456$\lambda$ | ~0.11 | 254.1 | 100.8 | 479.6 | 15.8 |
| 10 | ∞ | $+R_2$ | $+R_3$ | ∞ | 20.0 | 1.732$\lambda$ | ~0.06 | 302.7 | 121.6 | 577.3 | -0.4 |
| 11 | $-R_1$ | ∞ | $+R_3$ | ∞ | 19.9 | 2.888$\lambda$ | ~0.04 | 493.6 | 196.0 | 938.9 | 4.6 |

Versions #2, #3b and #8 can be used if enough accurate knowledge about the glass parameters are available and the calculated radii are kept within tight tolerance. However, the best way is to reduce a relative aperture of achromatic doublet which was designed with help of these versions. Herewith, the relative aperture should be selected smaller than 1/22 and this requirement applies at ideal spherical surfaces. If we take into account that real surfaces always have non-perfect shape then the relative aperture should be reduced additionally. Otherwise, many steps of the trial-and-error method will be required for reducing spherical aberration and as result an economic efficiency will be lost. Many labor hours should be spent on like type of achromatic doublet at poor predictability for their outcome. At the same time they have more losses of light (contrast) due to the additional flint lens in compare with standard non-achromatic object-lens at comparable light gathering (relative aperture). Therefore, it is quite expected this design of object-lens did not cause enthusiasm among opticians in middle 50th of XVIII century.

[7] Here in the table:
$R_1$, $R_2$ – first and second radii (by ray path) of the flint lens; $R_3$, $R_4$ – the similar radii for the crown lens. The signs before the radii are selected in accordance with modern rules applying at description of the optical systems;
F/D – the focal ratio obtained in Zemax's calculations takes into account the thickness and air gap (if necessary) between lenses;
GEO, RMS – the envelope and RMS radii respectively on the spot diagram (for reference, the Airy radius is about 13μm);
TSPH, TAXC –spherical and axial chromatic aberration coefficients respectively, letter 'T' means the transverse. The negative sign for spherical and axial chromatic aberrations indicates over-correction.
Peak-to-Valley and Strehl Ratio are calculated for the focal plane, where is minimum RMS of wave front error.

The version #2 was described by Benjamin Martin in 1759:

> *"…if we take any double and equally convex lens of crown glass, and another double concave of white flint,.. the radius of the internal cavity is the same as the radius of convexity in the [crown] lens… Hence appears the reason why this [flint] lens should be a double concave of unequal radii: and we see the true form of the compound object-glass in a telescope made on this principle, and constructed according to the theory, and which is different from any Thing we have yet seen made."*[8]

Just below, he described his opinion about this version of the achromatic doublet and claimed that given version before has been no made. There he also claimed that an achromatic lens "*…consisting of a double convex lens and plano-concave with its plain side outwards"* (version #1) has been made[9].

The version #3b with mediocre quality is different from the version #3 only orientation of the crown lens. So, it should be considered as incorrectly assembled the version #3.

Only four flint-forward versions #1, #3, #4 and #5 can be confidently applied at the focal ratio 20 as the initial approximation for achromatic doublets even if their surfaces will have imperfections.

The version #4 is best between them. This design was introduced by Alexis Clairaut in 1761 and was described him in the postscript of letter to Gerhard Müller[10] (Imperial Academy of Sciences and Arts, St. Petersburg). The first achromatic telescope according to the Clairaut's principles was made by de Létang to end of July 1761.

Distinctive features of the version #5 it has stronger radii (about 1.5 time) on the contiguous surfaces in compare with versions #1 and #3 and the flint lens has a meniscus profile as in Clairaut's design. Probably this peculiarities of version #5 pushed away from it of opticians in XVIII century. They were aware about the potential issues coupling with manufacturing a stronger radii and lens with meniscus profile. Clairaut complained in a letter to Daniel Bernoulli: *"…I am afraid that for a long time these telescopes [his design] will not become commonplace due to difficulties at their manufacture"* and to Leonard Euler that *"The dimensions [radii], that I give a two contiguous lenses which forming object-lens, are different from those in Dollond and definitely [object-lens is] better for the theory, and it seemed to me [it should be] better for practice too, but actually more difficult to perform well"*[11]. Herewith, the radii in Clairaut's design are significantly weaker in compare with version #5. The first reliable confirmation of use of the version #5 refers to second half of the XIX century. It is Steinheil flint-forward achromatic doublet[12]. Although, this version was described by Clairaut in his "Memory" which was published in 1764[13].

According to the article [1] the version #1 is attributed to James Ayscough and Joseph Linnell. While the version #3 was used by J. Dollond & Son [2]. It is known that Ayscough accomplished an orders on manufacturing of achromatic doublets for Chester Moor Hall in beginning 50th of XVIII century. In this regard, there are several questions:

- ? If in the versions #1 and #3 the original spherical aberrations are comparable and both they have significantly smaller aberrations than in version #8. Why then version #1 didn't widely used before 1758 year instead of version #8?

---

[8] B. Martin, *New Elements of Optics*, (1759), notes: 263-266, pp. 77-79 and Plate IV, Fig. VIII.

[9] *Ibid*, notes: 288-290, pp. 84-85.

[10] Reference [5], note 148, pp. 314-315.

[11] *Ibid*, note 164, p.318.

[12] *C.A. Steinheil & Söhne* optical company was established in 1855 in Munich. If to merge by the flat surfaces of the crown lenses of flint-forward and crown-forward doublets (both versions #5) in one triplet, then we will get very well-known Steinheil magnifier. If to place the diaphragm between two the same separated doublets as before then we will get Steinheil Aplanat or Dallmeyer Rapid-Rectilinear photography lens (invention of 1866 year).

[13] A. Clairaut, *Troisième mémoire sur les moyens de perfectionner les lunettes d'approche, par l'usage d'objectifs composés de plusieurs matières différemment réfringentes,* Histoire de l'Académie Royale des Sciences, 1762 (1764), p. 620.

- ? Why William Eastland in his testimony described version #8 instead of the better version #1? Though in beginning 1750s he was subcontractor for James Ayscough and he got the knowledge from him how to make an achromatic doublets.
- ? Why John Dollond did not care about violation of the patent (Eastland testimony), but on the other hand, he proposed to bring an action against the said Eastland for making and vending the achromatic telescopes (Watkins testimony)[14]?

The most likely answer to these questions is that the versions #1 and #3 emerged practically simultaneously in 1758, but version #1 had to appear a little later, maybe in 1759. And none of the versions cannot be attributed directly to Chester Moor Hall authorship. The following recognized events of 1758 push to such answer:

- ✓ Chester Moor Hall visited to Dollond's house in soon after and in connection with the issuance of the patent[15]; just at that moment John Dollond received the full information about when and how he came to the achromatic doublet; what method he used for calculation (apparently version #8) and who did the object glasses for him (Ayscough and Eastland).
- ✓ John Dollond visited to Ayscough's optical shop where he had conversation with James Ayscough and William Eastland[16]. There he let them knew about the patent. In this moment Dollond personally convinced these manufacturers of the achromatic doublets by the Moor Hall method cannot compete with him.
- ✓ A buoyant demand on new achromatic telescopes there was in all European market after middle of 1758 and they formed a serious competition for mirror telescopes.

Very likely, already to the end of 1758, James Ayscough and William Eastland realized that they were not able to compete in the production of achromats with Dollond and Watkins workshops. They had two ways, either to invent their own a new method of making doublets, or to use the Dollond's method as more perspective. Apparently the second way also was used (most likely by Eastland[17]) that eventually caused to the Dollond's discontent.

In reality, the invention of new method instead of Dollond's method is not so difficult if to understand that:

> *"…the removal of one impediment [chromatic aberration] had introduced another equally detrimental (the same I had before found in two glasses with water between them): for the two glasses, that were to be combined together, were segments of very deep spheres; and therefore the aberrations from the spherical surfaces became very considerable, and greatly disturbed the distinctness of the image… I plainly saw the possibility of making the aberrations of any two glasses equal; and as in this case the refractions of the two glasses were contrary to each other, their aberrations, being equal, would entirely vanish"*[18].

It becomes obvious how to change the parent version #8 to reduce the original spherical aberration in it. The first thing that comes in mind to distribute evenly the optical power on two surfaces in the flint and/or the crown lenses. If the distribution of the optical powers will be made for the both lenses, then we immediately get version #0, or version #1 if only on the crown lens, or version #7 if only on the flint lens. With a little efforts any optician could easily find from these three variants only one is successful.[19] This is version #1. Since the versions #1 and #3 have a pretty good and comparable quality of approximation. Why then during the court trials did not reveal a noticeable number of achromates which produced before 1758? It will remains unclear until we will not suppose that version #1 was realized a little later than version #3.

On the other hand, very well known that Joseph Linnell was free of the Spectaclemakers Company after June 30th, 1763, therefore the telescope signed by him was made after this date. The telescopes signed by Ayscough could be made during in the last life year of James Ayscough (d.1759), or possibly in the time when his widow Marta managed the workshop (until 1764). At this moment, there are no unambiguous evidences that Ayscough's and Linnell's telescopes, which were described in article [1], were produced before the middle of 1758.

---

[14] Reference [3], pp.293-294.

[15] The British patent #721: "A ***new method*** of making the objective glasses of refracting telescopes by compounding mediums of different refractive qualities, whereby the errors arising from the different refrangibility of light, as well as those which are produced by the spherical surfaces of the glasses, are perfectly corrected."

[16] At this meeting, James Champney and Christopher Stedman were also present. Reference [4], p.147

[17] Reference [5], pp.299-300.

[18] J. Dollond, *An Account of Some Experiments Concerning the Different Refrangibility of Light.*, Phil. Trans., vol.50 (1758), pp.741-742.

[19] Indeed, optician could find also another successful variant. It is version #7, but already with the crown lens in front.

**Crown-forward achromatic doublets**

Possible versions of the crown-forward doublets are shown on Figures 4 and 5. The formulas for calculation of the radii for each versions are collected in Table I.2, in Appendix I. As is before for the versions of flint-forward type the variants with only a numeric label are initial. All other versions on Figure 4 differ from the initial (numbered) variants by reverse orientation of the flint and/or crown lenses, and the absolute values of radii are the same as in corresponding initial variants.

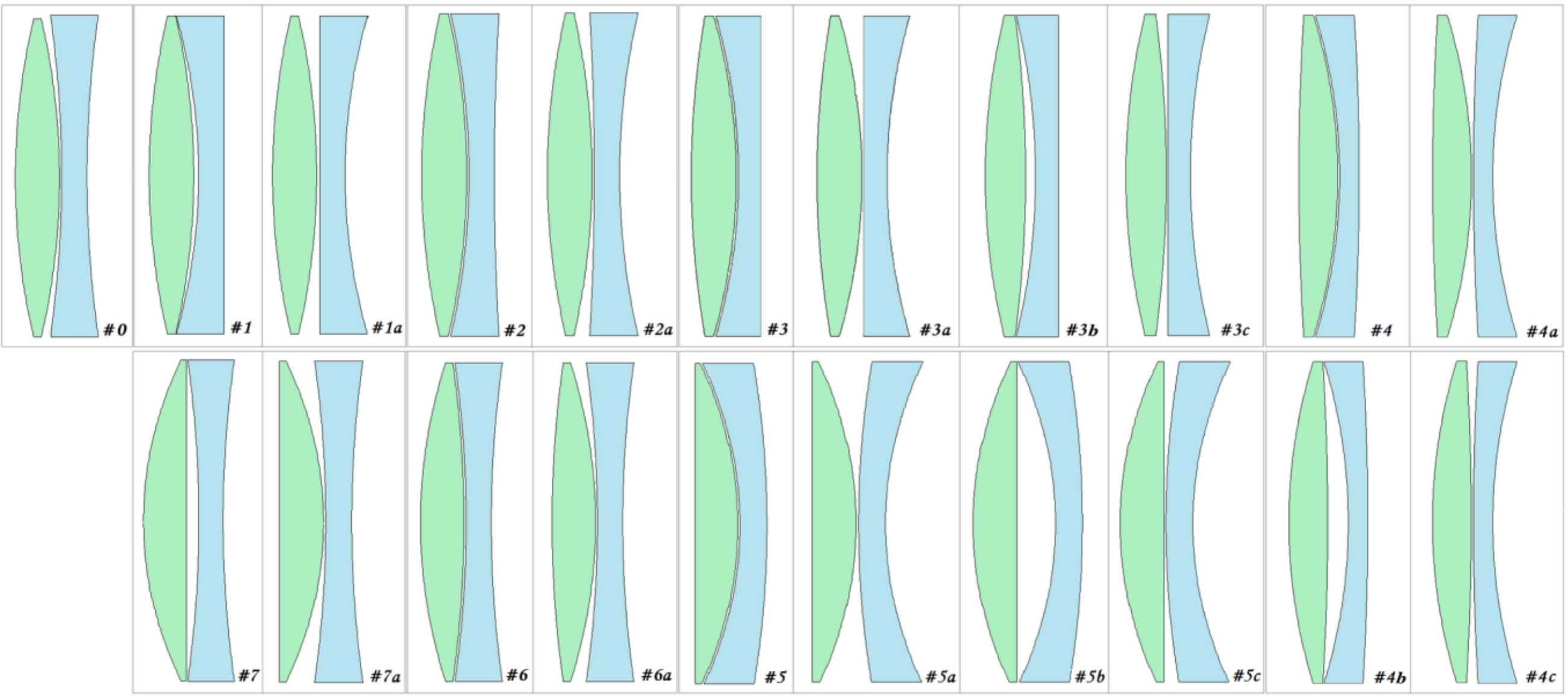

Figure 4. Some approximation versions of the crown-forward achromatic doublets are shown. The front green part is the crown lens. The versions with additional letter in label have the same radii as in initial versions with corresponding number, only orientation of a flint and/or crown lenses are different.

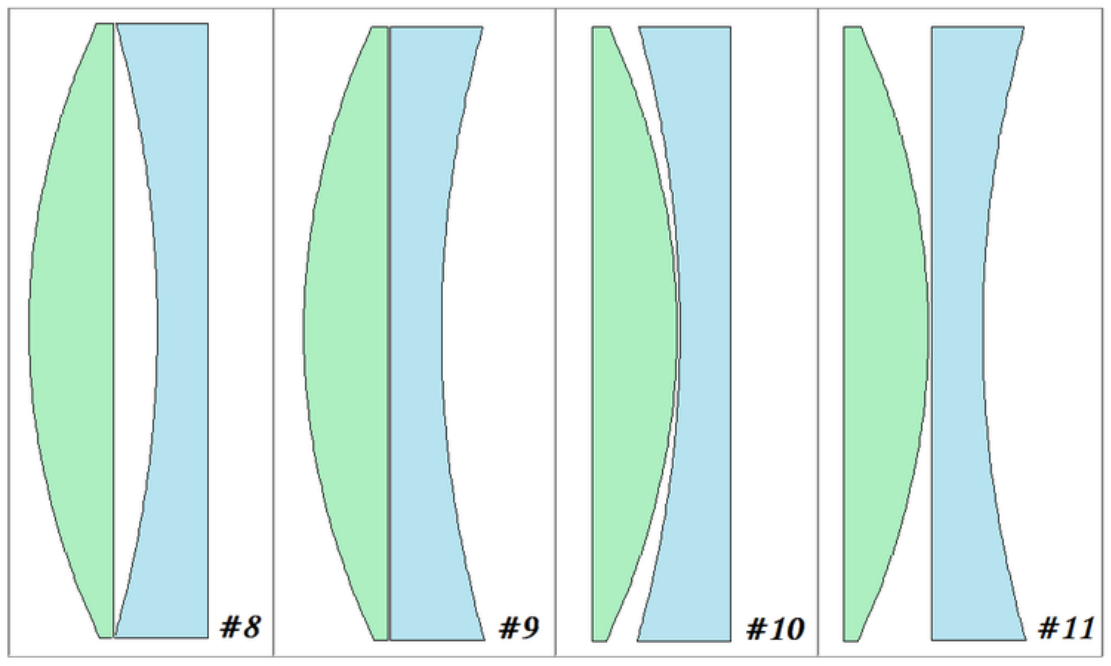

Figure 5. The versions of crown-forward doublets by Chester Moor Hall authorship.

These four versions on Figure 5 should also be considered as corresponding to the method of making achromatic doublets which was described in the testimony of William Eastland in the court, because the lens sequence in this achromatic doublet was not recorded. According to the results from Table 2 all these versions have big initial spherical aberration, more than 0.8λ. So, there is only one possibility for the achromatic doublet – this is version #8 and only in the flint-forward sequence. Perhaps William Eastland knew about this feature of the said doublet. Therefore, he did not clarify the lens sequence in the court. It was so obvious for him. The same words can be said about version #1 which could be attributed to James Ayscough and Joseph Linnell. In this case the crown-forward variant is not acceptable for practical use. So, we have to attribute the version #1 exclusively to the flint-forward type. Based on these two facts, we can argue that invented first achromatic doublet was flint-forward type. No doubt Chester Moor Hall could have in his possession exclusively the flint-forward achromatic object-lenses.

As was said before the version #2 with a flint lens in front (attributed to Benjamin Martin) has mediocre quality. However, the same version in the crown-forward variant has very small original spherical aberration.

Even if the optician designed the flint-forward doublet using this version after this object-lens will be made, he must have a desire to change the lens sequence on to the crown-forward variant.
In present time, the crown-forward version #2 is attributed to Joseph von Littrow, i.e. this invention of XIX century. Nevertheless, it is very hard to believe that Benjamin Martin did not make attempt to change the lenses sequence in the doublet in XVIII century. Apparently this design of an achromatic doublet should be attributed to Martin with the same right.

Table 2. Comparison of the initial approximations for the crown-forward achromatic doublets[20]

| # | $R_1$ | $R_2$ | $R_3$ | $R_4$ | F/D | Peak-to-Valley (λ=0.546μm) | Strehl Ratio | GEO Radius (μm) | RMS Radius (μm) | TSPH (μm) | TAXC (μm) |
|---|---|---|---|---|---|---|---|---|---|---|---|
| 0 | $+R_1$ | $-R_1$ | $-R_3$ | $+R_3$ | 20.0 | 0.677λ | 0.204 | 122.6 | 48.4 | 230.7 | 5.5 |
| 1 | $+R$ | $-R$ | $-R_3$ | $\infty$ | 20.0 | 0.570λ | 0.329 | 101.1 | 41.1 | -190.6 | 5.8 |
| 1a | $+R$ | $-R$ | $\infty$ | $+R_3$ | 19.8 | 1.314λ | ~0.13 | 237.6 | 93.2 | 446.3 | 9.9 |
| 2 | $+R$ | $-R$ | $-R$ | $+R_4$ | 20.0 | 0.030λ | 0.997 | 6.1 | 2.2 | 10.7 | 4.0 |
| 2a | $+R$ | $-R$ | $-R_4$ | $+R$ | 19.7 | 1.118λ | ~0.11 | 201.9 | 79.5 | 380.7 | 7.8 |
| 3 | $+R_1$ | $-R$ | $-R$ | $\infty$ | 20.1 | 0.048λ | 0.992 | 9.1 | 3.5 | -16.5 | 1.8 |
| 3a | $+R_1$ | $-R$ | $\infty$ | $+R$ | 19.9 | 1.839λ | ~0.05 | 330.0 | 130.4 | 625.5 | 8.3 |
| 3b | $+R$ | $-R_2$ | $-R$ | $\infty$ | 19.7 | 0.928λ | ~0.09 | 161.2 | 65.1 | -313.5 | 10.5 |
| 3c | $+R$ | $-R_2$ | $\infty$ | $+R$ | 19.8 | 0.953λ | ~0.09 | 173.6 | 67.3 | 322.2 | 12.1 |
| 4 | $+R_1$ | $+R_2$ | $+R_2$ | $-R_1$ | 20.0 | 0.078λ | 0.980 | 15.3 | 5.5 | -26.7 | 0.2 |
| 4a | $+R_1$ | $+R_2$ | $+R_1$ | $-R_2$ | 19.7 | 2.993λ | ~0.04 | 515.7 | 201.6 | 1009 | 10.5 |
| 4b | $-R_2$ | $-R_1$ | $+R_2$ | $-R_1$ | 19.4 | 2.142λ | ~0.06 | 373.8 | 150.4 | -708.3 | 20.7 |
| 4c | $-R_2$ | $-R_1$ | $+R_1$ | $-R_2$ | 19.4 | 0.832λ | ~0.10 | 150.1 | 55.9 | 278.2 | 19.5 |
| 5 | $\infty$ | $-R$ | $-R$ | $-R_4$ | 20.0 | 0.014λ | 0.999 | 4.7 | 1.1 | -3.6 | 0.5 |
| 5a | $\infty$ | $-R$ | $+R_4$ | $+R$ | 19.5 | 4.259λ | ~0.03 | 726.3 | 283.4 | 1412 | 14.5 |
| 5b | $+R$ | $\infty$ | $-R$ | $-R_4$ | 19.1 | 3.263λ | ~0.04 | 567.0 | 227.8 | -1041 | 32.1 |
| 5c | $+R$ | $\infty$ | $+R_4$ | $+R$ | 19.1 | 0.811λ | ~0.11 | 148.6 | 53.9 | 265.6 | 28.3 |
| 6 | $+R_1$ | $-R$ | $-R$ | $+R$ | 19.7 | 0.222λ | 0.847 | 40.7 | 15.2 | 76.3 | 8.8 |
| 6a | $+R$ | $-R_1$ | $-R$ | $+R$ | 19.9 | 1.460λ | ~0.12 | 254.9 | 100.6 | 500.7 | 3.3 |
| 7 | $+R_1$ | $\infty$ | $-R$ | $+R$ | 19.3 | 0.139λ | 0.939 | 28.4 | 9.3 | 48.5 | 21.8 |
| 7a | $\infty$ | $-R_1$ | $-R$ | $+R$ | 19.9 | 3.565λ | ~0.04 | 611.5 | 242.2 | 1215 | 1.8 |
| 8 | $+R_1$ | $\infty$ | $-R_3$ | $\infty$ | 19.2 | 1.090λ | ~0.11 | 189.4 | 75.4 | -355.3 | 25.3 |
| 9 | $+R_1$ | $\infty$ | $\infty$ | $+R_4$ | 19.3 | 0.786λ | ~0.12 | 139.0 | 52.5 | 262.6 | 20.6 |
| 10 | $\infty$ | $-R_2$ | $-R_3$ | $\infty$ | 20.0 | 2.280λ | ~0.07 | 397.2 | 158.0 | 777.0 | -0.3 |
| 11 | $\infty$ | $-R_2$ | $\infty$ | $+R_4$ | 19.8 | 4.232λ | ~0.03 | 719.8 | 283.7 | 1431 | 6.3 |

So, the version #2 should be attributed to the crown-forward type especially if the relative apertures of achromatic doublet will be faster than 1/20.
No doubt that versions #6 and #7 can exclusively be used for calculation of the crown-forward doublets. At the same time they have acceptable but a worse quality in comparison with versions #2, #3, #4 and #5. Therefore, achromatic doublets which were made based on these version should be rare.
The crown-forward versions #3 (attributed to Dollond), #4 (attributed to Clairaut) and #5 (attributed to Steinheil) have excellent correction of the aberrations. They are universal initial approximations for the both types of achromatic doublets. Visible differences in aberrations are negligible in practice. A more important factor is stability of a different glasses in aggressive environment. Crown glass is more stable than flint glass, therefore crown-forward doublets finally supplanted corresponding to them flint-forward variants.

[20] Here in the table:
$R_1$, $R_2$ – first and second radii (by ray path) of the crown lens; $R_3$, $R_4$ – the similar radii for the flint lens. Everything else are the same as in Table 1, see footnote 7.

**Summary**

The twenty five possible versions of flint-forward and the same quantity of crown-forward of the achromatic doublets in "thin-lens" approximation were considered. A qualitative comparative analysis of the versions by values of original (spherical and chromatic) aberrations for the optical systems with non-zero thickness of the lenses and gaps between them (if it was required) was carried out. Herewith the radii of the lens surfaces had the same values which were obtained from the formulas for "thin-lens" approximation. This approach allows to evaluate the quality of a particular approximation version and its applicability for making of the real achromatic doublets.
Among four the flint-forward and the same quantity of the crown-forward versions which undoubtedly should be attributed to Chester Moor Hall according to the testimony of William Eastland exists only one acceptable. This is the flint-forward version #8. Herewith, this version has mediocre quality due to relatively big original spherical aberration. Therefore it should be used at the relative aperture smaller than 1/22 (at ideal spherical surfaces), and with considering imperfections of surfaces it must be even less.
According to the letter signed by Veritas[21], Chester Moor Hall ordered lenses for the first achromatic doublet with relative aperture about 1/8 at two opticians (Edward Scarlett and James Mann Jr.) but both have engaged one and the same optician (George Bass). In this curious case, it is impossible to obtain a working achromatic doublet due to an expected differences in glass parameters (especially for flint) in the prism-experiments and in the real doublet. And especially with such inadequately large relative aperture for the used version #8. Obviously at that time Chester Moor Hall quite did not think about problems with a geometrical (mainly spherical) aberrations. He focused all his attention on achromaticity of the doublet. So, we have to admit that with high probability in the beginning 1730s Chester Moor Hall got a fail result – quasi-achromatic doublet with a huge spherical aberration even in assumption that George Bass made both lenses with perfect spherical and flat surfaces.
Apparently, the first achromatic doublet with acceptable correction of chromatic and spherical aberrations was made in Ayscough's workshop in first half of 1750s. At this time Chester Moor Hall no longer tried to keep his invention "in scrutoire" and the relative aperture for the doublet was selected much less in compare with the design of 1730s years. So, we come to next conclusions:

- Chester Moor Hall was inventor of the achromaticity in a glass doublet;
- Also he was one of inventors of the first working achromatic doublet, apparently together with James Ayscough.

Since after Ayscough was involved to resolve of the problems at the first doublet a real working achromatic lens was produced. Nevertheless, efforts at its manufacturing were so great and the differences in visible picture from non-achromatic lenses with comparable parameters were insignificant. Of course, if we suppose that the trial-and error method was not widely used due to limitation on a spent labor hours. So, this invention was not widely known until middle of 1750s (possibly even in London).
As in case with version #8 (attributed to Chester Moor Hall) the version #1 which attributed to the Ayscough's and after 1764 Linnell's workshop is suitable for practice only in flint-forward variant. Therefore, the first achromatic doublets which attributed directly or indirectly to Chester Moor Hall were flint-forward type. It is easy to see that the flint-forward versions #1 and #8 are different only by the shape of crown lens. Instead of the flat-convex crown lens in version #8 the version #1 has the double-convex lens with equal radii. At keeping of the optical power of crown lens (it is necessary for saving of achromaticity) the surface radii in the version #1 should be in two time weaker. Therefore, expected spherical aberration should be smaller. The calculation of real flint-forward doublet based on the version #1 proves that the given version has much less original aberrations in compare with the parent version #8. Furthermore, this version

[21] Gentleman's Magazine and Historical Chronicle, September 30th 1790, p.890: *"About 1733 he [CMH] completed several achromatic object-glasses that bore an aperture of more than 2½ inches, though the focal length did not exceed 20 inches."*

#1 has comparable quality with the flint-forward version #3 (attributed to the Dollond and Watkins workshops) and it is competitive to the last one. However, the court trials related to the Dollond's patent infringement revealed that an achromatic doublets were not widely distributed on a market before middle of 1758. Thereby they confirmed that the flint-forward version #1 should be appeared later of the flint-forward version #3.

In 1759 Benjamin Martin in his book (footnotes 8 and 9) informed about the flint-forward version #1 where he claimed that this variant of an achromatic doublet already has been made. In the same book he introduced another own version of the flint-forward doublet (version #2) which before has been no made. At moment of the book issue Benjamin Martin did not know that this design is better to use in the reverse (crown-forward) lens sequence than described by him. So, this means he did not pay attention on a spherical aberration and he had no a theory for calculation of aberrations in doublets (at least at that time).

The version #3 is universal and it can be used in practice in both variants. Most likely it was invented by John Dollond in beginning 1758 in flint-forward variant in like manner as the first achromatic doublet of Chester Moor Hall. This version was a basis for the patent #721 (footnote 15). Apparently J. Dollond & Son knew they were not the first in this invention[22]. Therefore, for prevent any possible complications with obtaining of the patent they decided to request it on the '*new method of making*' instead of the invention as such, that could be expected after reading of the Dollond's account (footnote 18) in Royal Society.

Under the "method of making" here we have to understand[23]:

- ✓ Formulas for the focal length of doublet and the achromatic condition at the selected glass pair;
- ✓ The particular version for volitional selection of the radii.

The further events have shown that the patent has protected the Dollond's new method from unfair replication.

The versions #4 (flint-forward) and #5 (crown-forward) are the best from theoretical point of view. They have minimal spherical aberration between all investigated versions. Also both versions have a flint lens with meniscus profile and strong radii (especially at version #5) on contiguous surfaces in compare with versions #1, #2 and #3. These features led to difficulties in production of the object-lenses in the XVIII century. The first object-lens according to version #5 was released in Steinheil's optical company only in the XIX century. And the first object-lens by version #4 (introduced by Clairaut) was dated by middle of 1761. However, according to Clairaut it was produced slowly and with significant difficulty. Therefore, the versions #1 (flint-forward), #2 (crown-forward) and #3 (in both variants) are the best in a practice in second half of the XVIII century.

And finally need to specify, many opticians in XVIII century preferred to change of flat surface on lenses to spherical surface with largest radius available in optical shop. The reason was that spherical surface with optical quality can be performed easily in comparison with flat surface the same quality. Apparently, such correction in the primary design was carried out to eliminate unpredictable deformation of the flat surface due to partial removal of internal (residual) stresses in glass and Twyman effect, which occur every time after making concave or convex surfaces on the opposite side of a lens. The changes in the initial approximation can be calculated easily before manufacturing stage (for details, see in Appendix I). Therefore, the trial-and-error method was not really required. In this case, it is important to choose the correct sign for the spherical surface to avoid growth of spherical aberration in the achromatic doublet.

---

[22] J. Dollond was informed about the new compound object glass by J. Ayscough, R. Rew and G. Bass. Reference [4], pp.118-122.

[23] Method of measurement of the glass parameters (refractive index and dispersion) should also be included in the "method of making", because the labor efficiency at this measurements was a critical part in a competitive struggle for the optical market.

**Appendix I**

Before proceeding to obtain formula for the focal length of lens with arbitrary radius of curvature, we have to take a few agreements:

1. if the light moves in optical system from left to right then all distances or thicknesses will be positive and they will be negative in opposite direction of movement;
2. any angles between two rays are positive if they will be counted from the first ray to the second in clockwise (CW) direction and they are negative if in counterclockwise (CCW) direction;
3. the radius is positive if the center of curvature lies on the right side from the vertex of surface and it is negative at the opposite position.

All four possible cases are shown on Figure I.1. Two cases (*a* and *b*) in the top row show a positive refraction or deflection of a ray to the optical axis, and the bottom row (*c* and *d*) represents the cases with negative refraction angles – $\theta$. The incidence angle – $\alpha$ is counted from the incident ray to the normal to surface.

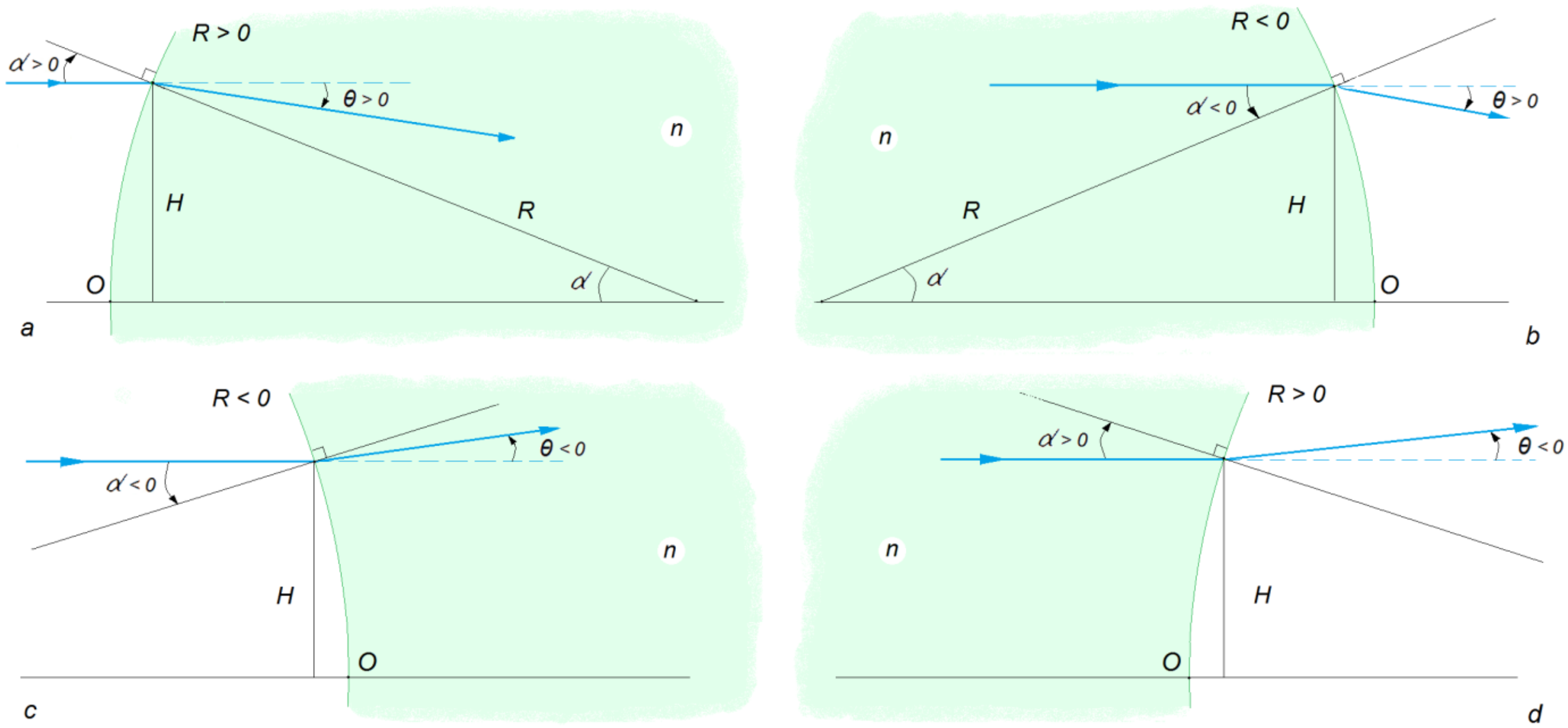

Figure I.1. The agreement about signs for angles and radii. The angle between two rays is positive when the first ray is rotated to the second in clockwise direction. It is negative in counterclockwise direction. The radius is positive when the center of curvature is placed on the right side from the vertex of surface. It is negative at the opposite position. It will always be assumed the light moves from left to right.

The law of refraction[24] establishes the relation between the angles of incidence and refraction. Here it should be noted that in the standard representation of the law under an angle of refraction, the angle between the refracted ray and the normal to surface is taken. In this article, the angle of refraction means the angle between the incident and refracted rays. This definition is better suited if we are considering the refractive power of a prism or lens.

The system of equations describes the propagation of an arbitrary ray through a lens:

$$\begin{cases} sin(\alpha_1) = n \cdot sin(\alpha_1 - \theta_1) \\ n \cdot sin(\alpha_2 - \theta_1) = sin(\alpha_2 - \theta_1 - \theta_2) \end{cases} \qquad (I.1)$$

---

[24] William Whewell, *History of the Inductive Science from the Earliest to the Present Times*, London: John H. Parker, 1837, p.276: *"The person who did discover the Law of the Sines, was Willebrord Snell, about 1621; but the law was first published by Descartes [Dioptrics, 1637], who had seen Snell's papers"*, according to Christiaan Huygens.

The angles $\alpha_1$, $\alpha_2$, and $\theta_1$, $\theta_2$ are shown on Figure I.2. Here $n$ – refractive index of a glass. Here it is assumed that the lens is placed in a medium with refractive index $n=1$.

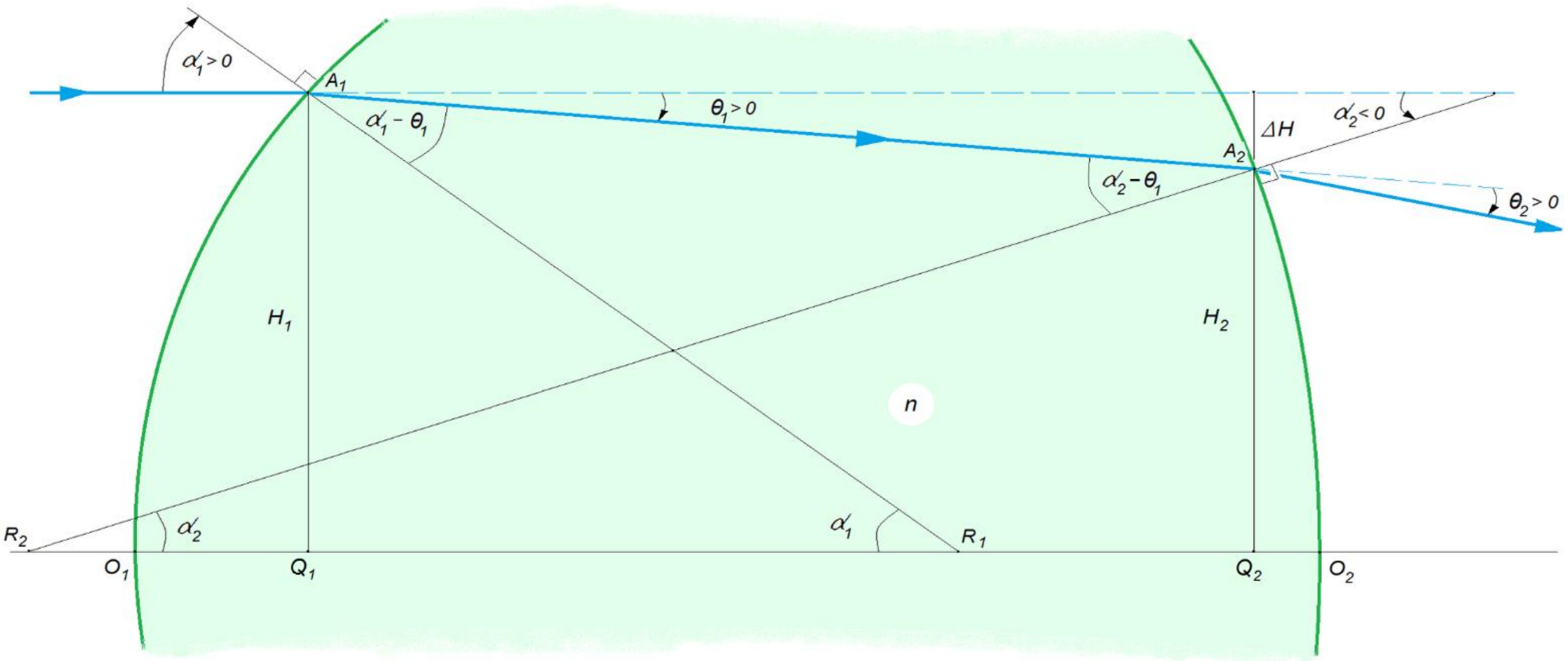

Figure I.2. Trajectory of a light (blue solid lines) through the thick focusing lens with curvature radii $R_1$, $R_2$. $A_1$ – input point and $A_2$ – output point on the lens which are placed on distance $H_1$ and $H_2$ respectively from the axis $O_1O_2$. The line $O_1O_2$ is the optical axis. The distance between points $O_1$ and $O_2$ is the central thickness ($t$) of lens.

Further, we assume that all angles $\alpha_{1,2};\ \theta_{1.2}$ are small enough. Therefore, the trigonometric functions of these angles can be replaced in (I.1) by angles themselves in the first-order expansion:

$$\begin{cases} \alpha_1 \approx n \cdot (\alpha_1 - \theta_1) \\ n \cdot (\alpha_2 - \theta_1) \approx \alpha_2 - \theta_1 - \theta_2 \end{cases} \qquad \text{(I. 2)}$$

The total refraction angle $\theta$ of lens is the sum of refraction angles $\theta_1$ and $\theta_2$. Excluding $n\theta_1$ from (I.2), we get the following expression for the total refraction angle:

$$\theta \equiv \theta_1 + \theta_2 \approx (n-1) \cdot (\alpha_1 - \alpha_2) \qquad \text{(I. 3)}$$

According to Figure I.2, $sin(\alpha_{1,2}) = H_{1,2}/R_{1,2} \approx\ \alpha_{1,2}$; here $R_{1,2}$ – the radii or lengths of lines $A_1R_1$ and $A_2R_2$ respectively. For refraction angles we have $tan(\theta_{1.2}) = H_{1,2}/F_{1,2} \approx \theta_{1.2}$; here $F_{1,2}$ – the distances between points $Q_{1,2}$ and focal points (not shown) for first and second surfaces respectively. The height of output ray is $H_2 = H_1 - \Delta H$. The negative sign was chosen, because for the focusing surface $\Delta H > 0$ is accepted. In the first-order approximation of the used trigonometric functions the distance $Q_1Q_2 \approx O_1O_2 \equiv t$ is the central thickness of lens and therefore $\Delta H \approx t \cdot \theta_1$. After substitution of the expressions for angles $\alpha_{1,2};\ \theta_{1.2}$ and $\Delta H$ in (I.3) and applying the first approximate equation from the system (I.2), we obtain a well-known formula for optical power ($P$) or focal length ($F$) of the lens:

$$P = \frac{1}{F} \equiv \frac{1}{F_1} + \frac{1}{F_2} - \frac{t}{F_1F_2} \approx (n-1) \cdot \left[\frac{1}{R_1} - \frac{1}{R_2} + \frac{(n-1)}{n} \cdot \frac{t}{R_1R_2}\right] \qquad \text{(I. 4)}$$

Since the signs for the incoming variables were not used, this formula is correct for arbitrary form of the lens. As is in the "thin lens" approximation at retaining only of the first-order in the expansion of trigonometric functions, a spherical aberration does not arise, because the focal length does not depend on the height of incidence ray. Obviously, if we neglect the thickness ($t$) of lens in (I.4), we get the formula for "thin lens". Example of the relationship between an arbitrary lens and an equivalent prism in “thin-lens” approximation is shown on Figure I.3.

Well-known fact that the refractive index depends on a wavelength. In agreement with formulas (I.3) and (I.4), the optical power (or focal length) as well as the total refraction angle of a lens also depend on wavelength or they are chromatic values. Let's consider in “thin lens” approximation the difference in optical powers[25] at the blue ($\lambda_b$) and red ($\lambda_r$) edges of wavelength range:

$$\Delta P = P(\lambda_b) - P(\lambda_r) \approx [n(\lambda_b) - n(\lambda_r)] \cdot \left[\frac{1}{R_1} - \frac{1}{R_2}\right]$$

or

$$\Delta P \approx \frac{n(\lambda_b) - n(\lambda_r)}{n(\lambda_g) - 1} \cdot P(\lambda_g) \qquad \text{(I. 5)}$$

The similar consideration at the same assumptions gives the difference in the total refraction angles[26]:

$$\Delta\theta \approx \frac{n(\lambda_b) - n(\lambda_r)}{n(\lambda_g) - 1} \cdot \theta(\lambda_g) \qquad \text{(I. 6)}$$

Here $\lambda_g$ – the wavelength of green light near the middle of the blue-red range and the coefficient $\nu(\lambda_g) \equiv (n(\lambda_g) - 1)/(n(\lambda_b) - n(\lambda_r))$ is Abbe number for the wavelength $\lambda_g$.
The optical powers or the total refraction angles of flint (*f*) and crown (*c*) lenses in a doublet have a negative and positive signs respectively. Thus, they will compensate each other when they will put together. If the absolute values of dispersions for the flint and crown lenses will be equal then the doublet will be achromatic:
$|\Delta P_f| = |\Delta P_c|$ and $|\Delta\theta_f| = |\Delta\theta_c|$.

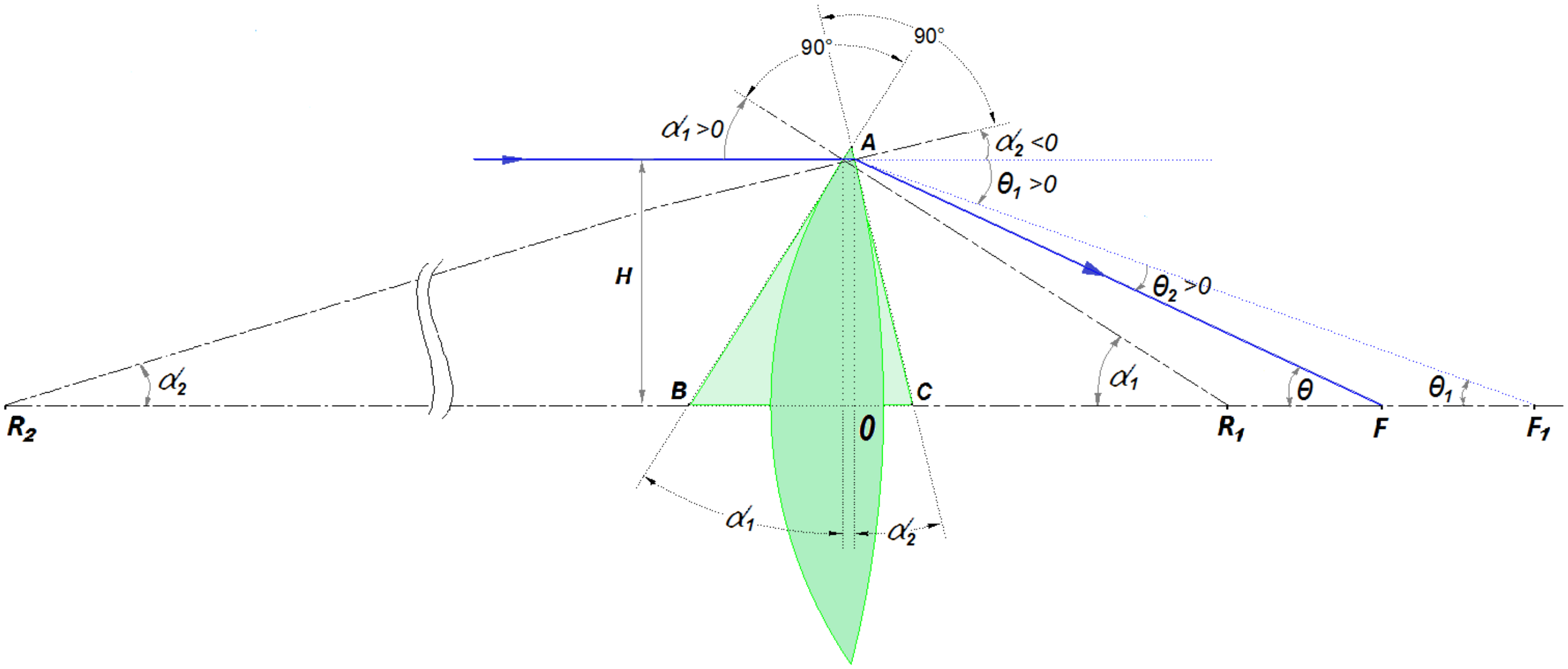

Figure I.3. The equivalent prism for the double-convex lens at the given beam height *H*. The blue solid lines are the ray trajectory through the lens or the equivalent prism *ABC*.
*F* – focal point of the lens or it is the focal length relative of point *O; $F_1$* – focal point of the first surface; $R_{1,2}$ – centers of curvature for the lens surfaces, herewith $R_1 > 0$ and $R_2 < 0$; $\theta_{1,2}$ – deflection angles relative of the initial direction for the first and second refractive surface respectively; in this case all angles are positive; $\alpha_{1,2}$ – apex angles of the equivalent half-prisms AOB and AOC respectively. The apex angle is unsigned. However the accident angle (equal to $\alpha_1$) between the incident ray and the normal is positive for the first surface and it is negative ($\alpha_2$) for the second surface.

[25] Sometimes this difference is called the dispersion of optical power.
[26] For the equivalent prism this value is called the angular dispersion.

Therefore, the achromatic condition for the doublet is:

$$\frac{P_c(\lambda_g)}{P_f(\lambda_g)} \approx \frac{\theta_c(\lambda_g)}{\theta_f(\lambda_g)} \approx -\frac{\nu_c(\lambda_g)}{\nu_f(\lambda_g)} \qquad \text{(I. 7)}$$

If we chose a focal length and a pair of glasses, which means the parameters $F;\ n_f;\ \nu_f;\ n_c;\ \nu_c$ are known, then the radii for flint and crown lenses for the different versions of achromatic doublets can be calculated in “thin lens” approximation with help the formulas from Table I.1 and Table I.2. The rows in the tables are sometimes divided by the double lines. This means that the radii between them should be calculated by the same formulas.
Suppose that the selected version has a lens with a flat surface and we would like to change it to a spherical surface with radius $R_0$. In this case, the achromaticity of doublet is violated. To restore achromaticity, it is necessary to change the optical power of another lens in the doublet. According to expression (I.7) ratio of the optical powers of crown and flint lenses in achromatic doublet should be equal to the ratio of their Abbe numbers. After making changes the ratio of the optical powers of the new lenses should be equal to the same ratio of the Abbe numbers because we use the same glass pair:

$$\frac{P_c(\lambda_g)}{P_f(\lambda_g)} \approx \frac{P_c(\lambda_g) + \Delta P_c}{P_f(\lambda_g) + \Delta P_f} \approx -\frac{\nu_c(\lambda_g)}{\nu_f(\lambda_g)}$$

Where $\Delta P_{f,c}$ – changes in the optical powers due to reshape of a flat surface and modification of another lens for compensation. It is easy to see that:

$$\frac{\Delta P_c}{\Delta P_f} \approx -\frac{\nu_c(\lambda_g)}{\nu_f(\lambda_g)} \qquad \text{(I. 8)}$$

As an example, consider the flint-forward version #1. According to Table I.1, the flint lens has first flat surface and second surface has a radius $R_2$, which is equal to:
$R_2 = \frac{1}{2}\frac{n_f - 1}{n_c - 1}\frac{\nu_c}{\nu_f} \cdot R$, here $R = 2(n_c - 1)\left[1 - \frac{\nu_f}{\nu_c}\right] \cdot F$ is radius of double and equally convex crown lens.

If we decide to change the flat surface on spherical with radius $R_0$ then the optical power of flint lens is changed on value: $\Delta P_f = (n_f - 1)/R_0$. We have two reasonable variants for modification of the crown lens: 1) to change both radii equally or 2) to change only the outer last radius. According to (I.8), in these two variants, the optical power of the crown lens should be changed on the values:

$$\Delta P_{c1} = 2(n_c - 1)\left[\frac{1}{R + \Delta R_1} - \frac{1}{R}\right] \approx -\frac{\nu_c}{\nu_f}\Delta P_f = -\frac{\nu_c}{\nu_f}\frac{(n_f - 1)}{R_0}$$
$$\Delta P_{c2} = (n_c - 1)\left[\frac{1}{R} + \frac{1}{R + \Delta R_2} - \frac{2}{R}\right] \approx -\frac{\nu_c}{\nu_f}\frac{(n_f - 1)}{R_0}$$

Where $\Delta R_1$ , $\Delta R_2$ – necessary changes of the radius $R$ in first and second variants respectively. Thus, we obtain dependencies on the variation in the radius of flint lens for both variants:

$$\Delta R_1(R_0) = R\frac{R_2}{R_0 - R_2} \quad \text{and} \quad \Delta R_2(R_0) = 2R\frac{R_2}{R_0 - 2R_2}$$

The variant with a symmetrical change in radii modifies the initial version less and half the value of the second variant is required. However, in some cases, it is preferable to keep the adjoined surfaces, especially with an equal radii and to modify only the outer surfaces, for instance, as in the versions #3 and #5.

Table I.1. The formulas for calculation of the radii for the flint-forward achromatic doublets.

| | $R_1$ | $R_2$ | $R_3$ | $R_4$ |
|---|---|---|---|---|
| #0 | $-R_2$ | $2(n_f-1)\left[\frac{\nu_c}{\nu_f}-1\right]\cdot F$ | $2(n_c-1)\left[1-\frac{\nu_f}{\nu_c}\right]\cdot F$ | $-R_3$ |
| #1 | $\infty$ | $\frac{1}{2}\frac{n_f-1}{n_c-1}\frac{\nu_c}{\nu_f}\cdot R$ | $R=2(n_c-1)\left[1-\frac{\nu_f}{\nu_c}\right]\cdot F$ | $-R$ |
| #1a | $-R_2$ | $\infty$ | $R$ | $-R$ |
| #2 | $\frac{-R}{2\frac{n_c-1}{n_f-1}\frac{\nu_f}{\nu_c}-1}$ | $R$ | $R=2(n_c-1)\left[1-\frac{\nu_f}{\nu_c}\right]\cdot F$ | $-R$ |
| #2a | $-R$ | $-R_1$ | $R$ | $-R$ |
| #3 | $\infty$ | $R=(n_f-1)\left[\frac{\nu_c}{\nu_f}-1\right]\cdot F$ | $R$ | $\frac{-R}{\frac{n_f-1}{n_c-1}\frac{\nu_c}{\nu_f}-1}$ |
| #3a | $-R$ | $\infty$ | $R$ | $R_4$ |
| #3b | $\infty$ | $R$ | $-R_4$ | $-R$ |
| #3c | $-R$ | $\infty$ | $-R_4$ | $-R$ |
| #4 | $2(n_f-1)\frac{\frac{\nu_c}{\nu_f}-1}{\frac{n_f-1}{n_c-1}\frac{\nu_c}{\nu_f}-1}\cdot F$ | $2(n_f-1)\frac{\frac{\nu_c}{\nu_f}-1}{\frac{n_f-1}{n_c-1}\frac{\nu_c}{\nu_f}+1}\cdot F$ | $R_2$ | $-R_1$ |
| #4a | $-R_2$ | $-R_1$ | $R_2$ | $-R_1$ |
| #4b | $R_1$ | $R_2$ | $R_1$ | $-R_2$ |
| #4c | $-R_2$ | $-R_1$ | $R_1$ | $-R_2$ |
| #5 | $\frac{R}{1-\frac{n_c-1}{n_f-1}\frac{\nu_f}{\nu_c}}$ | $R$ | $R=(n_c-1)\left[1-\frac{\nu_f}{\nu_c}\right]\cdot F$ | $\infty$ |
| #5a | $-R$ | $-R_1$ | $R$ | $\infty$ |
| #5b | $R_1$ | $R$ | $\infty$ | $-R$ |
| #5c | $-R$ | $-R_1$ | $\infty$ | $-R$ |
| #6 | $-R$ | $R=2\,(n_f-1)\left[\frac{\nu_c}{\nu_f}-1\right]\cdot F$ | $R$ | $\frac{-R}{2\frac{n_f-1}{n_c-1}\frac{\nu_c}{\nu_f}-1}$ |
| #6a | $-R$ | $R$ | $R_4$ | $-R$ |
| #7 | $-R$ | $R=2\,(n_f-1)\left[\frac{\nu_c}{\nu_f}-1\right]\cdot F$ | $\infty$ | $-\frac{1}{2}\frac{n_c-1}{n_f-1}\frac{\nu_f}{\nu_c}\cdot R$ |
| #7a | $-R$ | $R$ | $-R_4$ | $\infty$ |
| #8 | $\infty$ | $(n_f-1)\left[\frac{\nu_c}{\nu_f}-1\right]\cdot F$ | $\infty$ | $-(n_c-1)\left[1-\frac{\nu_f}{\nu_c}\right]\cdot F$ |
| #9 | $-R_2$ | $\infty$ | $\infty$ | $R_4$ |
| #10 | $\infty$ | $R_2$ | $-R_4$ | $\infty$ |
| #11 | $-R_2$ | $\infty$ | $-R_4$ | $\infty$ |

$R_1$, $R_2$ – the radii of the flint lens; $R_3$, $R_4$ – the radii of the crown lens. The rows are sometimes divided by the double lines. It means between them the radius values should be calculated by the same formulas. For example, in the row #1a first radius $R_1 = -R_2$, and $R_2$ equal to the value in accordance with the formula in the row #1 for appropriate radius.

Table I.2. The formulas for calculation of the radii for the crown-forward achromatic doublets.

| | $R_1$ | $R_2$ | $R_3$ | $R_4$ |
|---|---|---|---|---|
| #0 | $2(n_c-1)\left[1-\frac{\nu_f}{\nu_c}\right]\cdot F$ | $-R_1$ | $-2\left(n_f-1\right)\left[\frac{\nu_c}{\nu_f}-1\right]\cdot F$ | $-R_3$ |
| #1 | $R=2(n_c-1)\left[1-\frac{\nu_f}{\nu_c}\right]\cdot F$ | $-R$ | $-\frac{1}{2}\frac{n_f-1}{n_c-1}\frac{\nu_c}{\nu_f}\cdot R$ | $\infty$ |
| #1a | $R$ | $-R$ | $\infty$ | $-R_3$ |
| #2 | $R=2(n_c-1)\left[1-\frac{\nu_f}{\nu_c}\right]\cdot F$ | $-R$ | $-R$ | $\frac{R}{2\frac{n_c-1}{n_f-1}\frac{\nu_f}{\nu_c}-1}$ |
| #2a | $R$ | $-R$ | $-R_4$ | $R$ |
| #3 | $\frac{-R}{\frac{n_f-1}{n_c-1}\frac{\nu_c}{\nu_f}-1}$ | $R$ | $R=-(n_f-1)\left[\frac{\nu_c}{\nu_f}-1\right]\cdot F$ | $\infty$ |
| #3a | $R_1$ | $R$ | $\infty$ | $-R$ |
| #3b | $-R$ | $-R_1$ | $R$ | $\infty$ |
| #3c | $-R$ | $-R_1$ | $\infty$ | $-R$ |
| #4 | $2(n_c-1)\frac{1-\frac{\nu_f}{\nu_c}}{1-\frac{n_c-1}{n_f-1}\frac{\nu_f}{\nu_c}}\cdot F$ | $-2(n_c-1)\frac{1-\frac{\nu_f}{\nu_c}}{1+\frac{n_c-1}{n_f-1}\frac{\nu_f}{\nu_c}}\cdot F$ | $R_2$ | $-R_1$ |
| #4a | $R_1$ | $R_2$ | $R_1$ | $-R_2$ |
| #4b | $-R_2$ | $-R_1$ | $R_2$ | $-R_1$ |
| #4c | $-R_2$ | $-R_1$ | $R_1$ | $-R_2$ |
| #5 | $\infty$ | $R=-(n_c-1)\left[1-\frac{\nu_f}{\nu_c}\right]\cdot F$ | $R$ | $\frac{R}{1-\frac{n_c-1}{n_f-1}\frac{\nu_f}{\nu_c}}$ |
| #5a | $\infty$ | $R$ | $-R_4$ | $-R$ |
| #5b | $-R$ | $\infty$ | $R$ | $R_4$ |
| #5c | $-R$ | $\infty$ | $-R_4$ | $-R$ |
| #6 | $\frac{-R}{2\frac{n_f-1}{n_c-1}\frac{\nu_c}{\nu_f}-1}$ | $R$ | $R=-2\left(n_f-1\right)\left[\frac{\nu_c}{\nu_f}-1\right]\cdot F$ | $-R$ |
| #6a | $-R$ | $-R_1$ | $R$ | $-R$ |
| #7 | $-\frac{1}{2}\frac{n_c-1}{n_f-1}\frac{\nu_f}{\nu_c}\cdot R$ | $\infty$ | $R=-2\left(n_f-1\right)\left[\frac{\nu_c}{\nu_f}-1\right]\cdot F$ | $-R$ |
| #7a | $\infty$ | $-R_1$ | $R$ | $-R$ |
| #8 | $(n_c-1)\left[1-\frac{\nu_f}{\nu_c}\right]\cdot F$ | $\infty$ | $-\left(n_f-1\right)\left[\frac{\nu_c}{\nu_f}-1\right]\cdot F$ | $\infty$ |
| #9 | $R_1$ | $\infty$ | $\infty$ | $-R_3$ |
| #10 | $\infty$ | $-R_1$ | $R_3$ | $\infty$ |
| #11 | $\infty$ | $-R_1$ | $\infty$ | $-R_3$ |

$R_1$, $R_2$ – the radii of crown lens; $R_3$, $R_4$ – the radii of flint lens. All other conditions for the used formulas are the same as in case for Table I.1.